# Systematic Determination of Absolute Absorption Cross-section of Individual Carbon Nanotubes

Kaihui Liu*[1,2], Xiaoping Hong*[1], Sangkook Choi[1,3], Chenhao, Jin[1], Rodrigo B. Capaz[1,4], Jihoon Kim[1], Wenlong Wang[2], Xuedong Bai[2], Steven G. Louie[1,3], Enge Wang[5], Feng Wang[1,3]

[1] Department of Physics, University of California at Berkeley, Berkeley, CA 94720, USA

[2] Institute of Physics, Chinese Academy of Sciences, Beijing 100190, China

[3] Materials Science Division, Lawrence Berkeley National Laboratory, Berkeley, CA 94720, USA

[4] Instituto de Física, Universidade Federal do Rio de Janeiro, Caixa Postal 68528, Rio de Janeiro, RJ 21941-972, Brazil

[5] International Center for Quantum Materials, School of Physics, Peking University, Beijing 100871, China

* These two authors contribute equally to this work

**Abstract:**

**Optical absorption is the most fundamental optical property characterizing light-matter interactions in materials and can be most readily compared with theoretical predictions. However determination of optical absorption cross-section of individual nanostructures is experimentally challenging due to the small extinction signal using conventional transmission measurements. Recently, dramatic increase of optical contrast from individual carbon nanotubes has been successfully achieved with a polarization-based homodyne microscope, where the scattered light wave from the nanostructure interferes with the optimized reference signal (the reflected/transmitted light). Here we demonstrate high-sensitivity absorption spectroscopy for individual single-walled carbon nanotubes by combining the polarization-based homodyne technique with broadband supercontinuum excitation in transmission configuration. For the first time, high-throughput and quantitative determination of nanotube absorption cross-section over broad spectral range at single-tube level was performed for more than 50 individual chirality-defined single-walled nanotubes. Our data reveal chirality-dependent behaviours of exciton resonances in carbon nanotubes, where the exciton oscillator strength exhibits a universal scaling law with the nanotube diameter and the transition order. The exciton linewidth (characterizing the exciton lifetime) varies strongly in different nanotubes, and in average it increases linearly with the transition energy. In addition, we establish an empirical formula by extrapolating our data to predict the absorption cross-section spectrum for any given nanotube. The quantitative information of absorption cross-section in a broad spectral range and all nanotube species can not only provide new insight into the unique photophysics in one-dimensional carbon nanotubes, but also enable absolute determination of optical quantum efficiencies in important photoluminescence and photovoltaic processes.**

**Significance Statements**

Determination of optical absorption cross-section is always among the central importance of understanding a material. However its realization on individual nanostructures, such as carbon nanotubes, is experimentally challenging due to the small extinction signal using conventional transmission measurements. Here we develop a technique based on polarization manipulation to enhance the sensitivity of single-nanotube absorption spectroscopy by two-orders of magnitude. We systematically determine absorption cross-section over broad spectral range at single-tube level for more than 50 chirality-defined single-walled nanotubes. Our data reveals chirality-dependent one-dimensional photo-physics through the behaviours of exciton oscillator strength and lifetime. We also establish an empirical formula to predict absorption spectrum of any nanotube, which provides the foundation to determine quantum efficiencies in important photoluminescence and photovoltaic processes.

Single-walled carbon nanotubes (SWNTs), a model one-dimensional (1D) nanomaterial system, constitute a rich family of structures[1]. Each single-walled nanotube structure, uniquely defined by the chiral index (n,m), exhibits distinct electrical and optical properties[2-5]. Quantitative information of SWNT absorption cross-section is highly desirable for understanding nanotube electronic structures, for evaluating quantum efficiency of nanotube photoluminescence[5,6] and photocurrent[7-9], and for investigating the unique many-body effects in 1D systems[10-16]. Despite its obvious importance, reliable experimental determination of nanotube absorption cross-section at single-tube level is still challenging[17]. Previous absorption measurements on ensemble nanotube samples only provide averaged behaviour[18-20]. Recent absorption studies of individual nanotubes, suffering from small absorption signals and/or slow laser frequency scanning, cannot determine the absolute absorption cross-section and are limited in achievable spectral range[15, 21-23].

In this letter we demonstrate a high-sensitivity polarization-based homodyne method to measure nanotube absorption spectra. By manipulating the light polarization we enhanced the nanotube-induced transmission contrast, $\Delta I/I$, by two orders of magnitude, and this enhanced transmission contrast can be quantitatively related to nanotube absorption cross-section along and perpendicular to the nanotube axis. Using this polarization control together with supercontinuum laser source, we realized high-throughput and broadband absorption measurements at single-tube level. Combined with electron diffraction technique on the same tube, it enables absolute determination of absorption cross-sections of individual chirality-defined nanotubes for the first time. We obtained quantitative absorption spectra of over 50 SWNTs of different chiralities, and established a phenomenological formula for absorption cross-sections of different nanotubes. The chirality dependent nanotube absorption spectra reveal unique 1D photophysics in nanotubes, including a universal scaling behaviour of exciton oscillator strength and of exciton resonance linewidth.

## Results

**Experimental design.**

In two-dimensional (2D) monolayer graphene a universal absorption of ~ 2.3% in the visible and near infrared range was predicted and observed[24, 25]. For 1D nanotube with light polarized along its axis, the typical absorption of a single nanotube is 2 to 3 orders of magnitude smaller, i.e. at $10^{-4}$ to $10^{-5}$, due to its nanometer diameter compared to micrometer illumination beam size. This small signal is easily overwhelmed by intensity fluctuation of the light source, making it difficult to measure single-tube absorption with conventional transmission spectroscopy. However, realizing that conventional transmission measurement is just one form of homodyne detection, we can measure the nanotube absorption cross-section with high sensitivity by optimizing the homodyne process.

From the interferometric point of view, our technique is a special example of homodyne detection. This method has been recently extended to carbon nanotubes with manipulation of polarizations[26, 27]. In a homodyne measurement the detected light intensity change ($\Delta I/I$) originates from the interference between the signal ($E_s$) and local oscillator electric field ($E_{LO}$). Therefore

$$\frac{\Delta I}{I} = \frac{|E_{LO}+E_s|^2 - |E_{LO}|^2}{|E_{LO}|^2} = \frac{2|E_s|}{|E_{LO}|} \cos\phi \xrightarrow{\text{Transmission}} \frac{\Delta T}{T} = \frac{2|E_{NT}|}{|E_{in}|} \cos\phi = -\alpha, \qquad \text{(Eq. 1)}$$

where $\phi$ is the relative phase between $E_s$ and $E_{LO}$ and we have ignored the small $|E_s|^2$ term. The right equation describes conventional nanotube transmission measurements, where $E_s$ and $E_{LO}$ are respectively the nanotube forward scattering wave ($E_{NT}$) and the unperturbed incident light ($E_{in}$), and α is the nanotube absorption. If we can control $E_s$ and $E_{LO}$ with precisely defined relations to $E_{NT}$ and $E_{in}$, we can potentially great enhance optical detection contrast $\Delta I/I$ and from it determine the absolute nanotube absorption cross section. This can be achieved through polarization manipulation as shown in Fig. 1a. Two nearly-crossed polarizers (with a small deviation angle δ) were used to control the incident and outgoing light polarization, and two polarization-maintaining objectives were placed confocally between this polarizer pair. Suspended individual nanotubes were positioned at the focus of the objectives and with an angle of 45 degrees with respect to the first polarizer. This configuration varies both local oscillator $E_{LO}$ and signal $E_s$ in a precise matter, where local oscillator is greatly reduced by $E_{LO}=E_{in}\sin\delta$, and the signal is related to the initial nanotube field by $E_s \approx (E_{NT}^{//}-E_{NT}^{\perp})/\sqrt{2}$, where $E_{NT}^{//}$ and $E_{NT}^{\perp}$ are nanotube field along and perpendicular to the nanotube axis, respectively. With this polarization control, we can obtain the polarization-dependent nanotube absorption through the enhanced homodyne optical contrast directly using (Supplementary information S2)

$$-\frac{\Delta I}{I}(\delta) = \frac{\alpha_{//}-\alpha_{\perp}}{2}\text{ctg}(\delta) + \frac{\alpha_{//}+\alpha_{\perp}}{2}. \qquad \text{(Eq. 2)}$$

where $\alpha_{//}$ and $\alpha_{\perp}$ are nanotube absorption constants for light polarized parallel and perpendicular to the nanotube axis, respectively. Here $(\alpha_{//} - \alpha_{\perp})/2$ characterizes the strong depolarization effect of nanotubes. It gives rise to a greatly enhanced optical contrast $\Delta I/I$ at small δ, which can reach ~ 1% (compared to $10^{-4}$ in conventional transmission change) and becomes easily detectable.

**Determination of absolute absorption cross section.**

The power of our technique is demonstrated in Fig. 2, which displays $\Delta I/I$ spectra from a SWNT using different polarization settings. The chiral index of this nanotube was independently determined as (24,24) from its electron diffraction pattern (Fig 2a)[28]. By gradually decreasing δ (the deviation angle from the crossed polarizer position as labeled in Fig. 1a) from 5º to 0.5º, the resulting modulation signal $|\Delta I/I|$ increases steadily to 1% level at peak positions. Figure 2c further shows that when $\delta$ crosses zero, the local oscillator $E_{LO}$ changes sign, and so does the homodyne interference modulation $\Delta I/I$. This new polarization scheme, coupled with supercontinnum illumination and array detections, enables us to obtain single-nanotube $\Delta I/I$ spectra across a wide spectral range in a few seconds, orders of magnitude faster than previous approaches[15, 21, 22].

The spectra of $\Delta I/I$ as a function of $\delta$ allow direct determination of both $\alpha_{//}$ and $\alpha_{\perp}$. Figure S3 displays $\alpha_{//}$ and $\alpha_{\perp}$ over a broad spectral range for the (24,24) SWNT, which have values in the order of $10^{-4}$ as expected. Nanotube absorption cross-sections are proportional to $\alpha_{//}$ and $\alpha_{\perp}$, and their absolute values can be obtained once the illumination beam size is known. We determine the focused supercontinnum beam profile at high accuracy by systematically measuring the absorption spectra with nanotubes at different positions in the focused beam (Supplementary information S3). With this information, we obtained spectra of optical absorption cross-sections per atom along both parallel ($\sigma_{//}$) and perpendicular ($\sigma_{\perp}$) polarizations for the (24,24) SWNT, which is displayed in Fig. 2d. We performed such absorption measurements on 57 chirality identified SWNTs, and it allows us to systematically examine polarization dependent absorption cross-section in different nanotube species for the first time. The $\sigma_{//}$ spectra show prominent and different exciton transition peaks in different nanotubes. These chirality-dependent exciton transitions provide rich information on chirality-dependent nanotube photophysics, as we will describe later. The $\sigma_{\perp}$ spectra, on the other hand, show a small and finite perpendicular absorption ( ~ 1/5 of the average $\sigma_{//}$ value) and they are largely featureless.

We first examine briefly the $\sigma_\perp$ spectra, which have never been probed at single nanotube level before. The small magnitude of $\sigma_\perp$ can be attributed to depolarization effects[18, 19, 29, 30], but its lack of any resonance features is surprising at the first look. It is widely known that perpendicularly polarized light can excite $E_{\mu, \mu\pm1}$ transitions due to angular momentum selection rule, where $\mu$ is the sub-band index[1]. One naturally expects resonance features associated with $E_{\mu, \mu\pm1}$ exciton transitions in $\sigma_\perp$ spectra. Indeed prominent absorption peak corresponding to $S_{12}$ and $S_{21}$ transitions have been observed in semiconducting nanotubes with perpendicular polarization excitation[6]. However, here we do not observe any resonances in our $\sigma_\perp$ spectra, where transitions between higher sub-bands are probed. Detailed theoretical analysis shows that although $E_{\mu, \mu\pm1}$ transitions are symmetry allowed, their matrix elements are always zero close to the bandgap except for the $S_{12}$ and $S_{21}$ transitions[31]. This matrix element effect strongly suppresses exciton transition (as well as van Hove singularity at the bandedge), resulting in no spectral resonances for higher order $E_{\mu, \mu\pm1}$ transitions. Away from the bandedge, $E_{\mu, \mu\pm1}$ transition matrix element becomes finite and results in the finite but largely featureless $\sigma_\perp$, as we observe experimentally.

**Systematic analysis of absorption cross section**

Below we will focus on $\sigma_{//}$ spectra in chirality defined SWNTs. Figure 3a-d display four representative parallel-polarization spectra (black lines) in semiconducting (18,14), (25,23) and metallic (16,16), (24,24) nanotubes, where the chiral indices were independently determined by the electron diffraction techniques. All $\sigma_{//}$ spectra in Fig. 3 are characterized by sharp optical resonances arising from excitonic transitions[10-16] and a broad continuum background. They provide a wealth of information on the unique nanotube photophysics.

## Discussion

**Approximation sum rule.**

We first note that for all nanotubes, the average absorption cross-section is ~ $7.6 \times 10^{-18}$ cm$^2$/atom. This corresponds to an integrated cross-section ($\Sigma_A$) of $7.1 \times 10^{-18}$ eV·cm$^2$/atom between the spectral range 1.55 to 2.48 eV, the same as $\Sigma_A$ of graphene in the same range[24, 25]. Figure 4a shows in detail the distribution of $\Sigma_A$ in different nanotubes (dots), which converges on the graphene value (dashed line). The physical origin for this convergence is an approximated f-sum rule, where the integrated oscillator strength per atom over a sufficiently large spectral range should be the same for all graphitic structures. The agreement between nanotubes and graphene values provides an independent confirmation of the accuracy of our nanotube absorption cross-section results. (The slight variation in nanotube values is presumably due to the different distribution of resonant peaks in different nanotubes and the finite integrated energy range.)

**Chirality and transition dependent exciton oscillator strength.**

To quantitatively describe the chirality dependent absorption features in different nanotubes, we introduce a phenomenological model composed of a discrete exciton peak and corresponding continuum absorption for each subband. The exciton transitions are Lorentzian resonances, each characterized by its resonance energy $E_p$, resonance width $\Gamma_p$, and oscillator strength (integrated cross-section) $\Sigma_p$. Here p is an integer indexing optical transitions of both semiconducting ($S_{ii}$) and metallic nanotubes ($M_{ii}$) starting from 1 in the order of $S_{11}$, $S_{22}$, $M_{11}$, $S_{33}$, $S_{44}$, $M_{22}$, $S_{55}$, $S_{66}$, $M_{33}$, $S_{77}$,…[16]. The "continuum" absorption, including the contribution from band-to-band transitions as well as phonon sidebands and higher-order exciton transitions, is approximated empirically by a Lorentzian broadened form of the function $\Theta[E-(E_p+\Delta_p)] \times (1/\sqrt{E-(E_p+\Delta_p)})$.

Here $\Theta$ is the heavyside step function, $\Delta_p$ is the offset of continuum edge relative to the exciton transition, and the second term models the 1D density of states close to the band edge (Supplementary information S4). Red lines in Fig. 3 show that our phenomenological model nicely reproduces the experimental absorption spectra, with violet and dark yellow lines showing the exciton and continuum contribution, respectively. We note that the continuum absorption constitutes a significant portion of the total nanotube absorption oscillator strength, which has never been appreciated previously. The relative importance of continuum absorption increases with the nanotube diameter and the transition index p.

Exciton resonances, the most prominent features in the absorption spectra, encode rich 1D nanotube physics in their chirality dependent behaviour. (1) The exciton absorption oscillator strength provides new information on electron-hole interaction strength in the 1D nanotubes. (2) The exciton transition linewidth, arising from the finite exciton lifetime, reveals the ultrafast relaxation dynamics of excited states. (3) The exciton transition energies and their dependence on nanotube species observed in absorption spectra here provide a valuable confirmation of the recent assignment established using Rayleigh scattering spectroscopy[16] with an accuracy of 10 meV.

The exciton takes its oscillator strength from band-to-band transitions due to the electron-hole correlation[12-15]. Its value depends on both the interband optical transition matrix element and the exciton wavefunction at zero electron-hole separation. We observe in the experimental spectra (Fig. 3) that the exciton oscillator strength decreases significantly in large diameter nanotubes. A systematic examination reveals that the tube-dependent exciton oscillator strength can be described by a universal scaling law, $\Sigma_p \sim \dfrac{1}{(p+7.5)d}$, as displayed in Fig. 4b. For semiconducting nanotubes, this scaling law can be understood theoretically using a model description of the excitonic effects, and the magnitude of observed exciton oscillator strengths agrees well with our theoretical predictions[32].

Figure 4b also shows that the exciton oscillator strengths in metallic (diamonds) and semiconducting (circles) nanotubes have similar magnitude and fall on the same curve. This behaviour appears surprising, because the electron-hole interactions are expected to be much weaker in metallic nanotubes due to free electron screening. Naively the resulting exciton transition should have significantly smaller oscillator strength. Previous *ab initio* calculations[12,13], however, showed that the oscillator strength from excitons in metallic nanotubes are comparable to those in semiconducting tube, although no systematic analysis was carried out. Further effort will be required to understand quantitatively the experimental data here, and it could lead to deeper insight into many-body interactions in 1D nanotubes.

**Transition energy dependent exciton linewidth**

The optical linewidth of exciton transitions (for p>1) originates mainly from exciton lifetime broadening due to electron-electron and electron-phonon interactions. Ultrafast evolution of the exciton states plays a key role in important optoelectronic processes like multiexciton generation[9] and impact ionization[2]. Our data provide a unique opportunity to investigate the chirality dependent ultrafast dynamics in SWNTs. We plot in Fig. 4c the observed exciton linewidth in semiconducting (circles) and metallic (diamonds) nanotubes as a function of exciton transition energies. It shows that (a) the exciton linewidth, in average, increases linearly with transition energy and (b) there is a large variation of exciton lifetime in different nanotube species.

The linear increase of average linewidth indicates a shorter exciton lifetime sacles inversely with transition energy. This faster relaxation can be approximately attributed to increased phase space for electron-electron and electron-phonon scatterings at higher energy. Similar increase in excited state relaxation rate with energy has been observed in studies of graphite[33] and low-order transitions in carbon nanotubes[34]. The large variation of exciton lifetime, even for transitions at the same energy, indicates that the ultrafast electron relaxation may depend sensitively on the exact electronic structure, an important effect that has never been explored previously. We hope our new experimental data can stimulate their theoretical study to gain more insight on ultrafast electron dynamics in 1D systems.

**Empirical description of (n,m) nanotube absorption cross section**

Lastly we are able to develop an empirical formula for optical absorption spectra in all nanotubes (with diameter ranging from 1.2 nm to 3.2 nm) using the established relations for exciton oscillator strength and resonance width described above. This phenomenological description reproduces the absorption spectra within 20% accuracy in our experimental energy region 1.45-2.55 eV (See the detailed description and example in Supplementary information S4). Such empirical formula of optical absorption cross-section for a wide variety of nanotube across the visible spectral range will be a valuable reference for determining quantum efficiency of important optical processes like photoluminescence and photovoltaics.

**Methods**

(1) Sample preparation and chiral index (n,m) characterization

In this study we combined single-tube absorption spectroscopy and electron-diffraction techniques on the same individual suspended nanotubes. The experimental scheme is illustrated in Fig. S1. Suspended long nanotubes were grown by chemical vapor deposition (CVD) across open slit structures (~ 30×500 μm) fabricated on silicon substrates. We used methane in hydrogen ($CH_4:H_2$=1:2) as gas feedstock and a thin film (~ 0.2 nm) of Fe as the catalyst for CVD growth at 900 °C[35]. This growth condition yields extremely clean isolated nanotubes free of amorphous carbon and other adsorbates. We determined the atomic structure of every nanotube from the electron diffraction pattern using nano-focused 80 keV electron beams in a JEOL 2100 transmission electron microscope (TEM)[28]. By utilizing the slit edges as spatial markers, the same individual nanotubes can be identified in both TEM and optical microscope setup (Supplementary information S1).

(2) Measurement of the homodyne modulation $\Delta I/I$

We obtain the homodyne modulation spectrum by taking two sets of spectra with the nanotube in ($I_i$) and out ($I_o$) of the supercontinuum focus. The modulation signal is obtained as $\Delta I/I=(I_i-I_o)/I_o$. We use a supercontinuum laser as the light source and a spectrometer coupled with a linear array detector to record the spectra. By shifting the nanotube in and out of the focus quickly, we can take a high quality modulation spectrum within 2 seconds.

**Acknowledgement**: This study was supported by NSF CAREER grant (No. 0846648), NSF Center for Integrated Nanomechanical Systems (No. EEC-0832819), NSF Grant No. DMR10-1006184, DOE Contract No. DE-AC02-05CH11231, and DOE Molecular Foundry (No. DE-AC02-05CH11231); and by the 973 Project (Grant Nos. 2012CB933003, 2013CB932601, and 2013CB932603), NSFC (Grant Nos. 11027402, 91021007, 10974238, and 20973195) and CAS (Grant No.KJCX2-YW-W35) of China. Computational resources have been provided by NSF through TeraGrid resources at NICS and DOE at Lawrence Berkeley National Laboratory's NERSC facility.R.B.C. acknowledges support from Brazilian funding agencies CNPq, FAPERJ and INCT - Nanomateriais de Carbono.

**Figure captions**

**Figure 1. Scheme of polarization-optimized homodyne detection for single-nanotube absorption.** (**a**) Two nearly-crossed polarizers (with a small deviation angle δ) were used to control the incident and outgoing light polarization, and two polarization-maintaining objectives were placed confocally between this polarizer pair. Suspended individual nanotubes were positioned at the focus of the objectives and with an angle of π/4 with respect to the first polarizer polarization. We use broadband supercontinuum laser (spectral range 450-850 nm) as the light source and a spectrometer equipped with a silicon linear array for fast wide-spectral detection. (**b**) Interference between $E_{in}$ and $E_{NT}$ after the nanotube gives rise to nanotube extinction, while interference between $E_s$ and $E_{LO}$ after polarizer-2 generates the final homodyne signal. Given that $E_{LO}:E_{in} \sim 10^{-2}$ and $E_s:E_{NT} \sim 1/\sqrt{2}$ with suitable polarizer setting, the enhanced homodyne modulation signal ΔI/I can be two orders of magnitude higher than the transmission change ΔT/T=−α. It enables high sensitivity measurements of nanotube absorption α. The δ angle is exaggerated for better visualization.

**Figure 2. Representative data for polarization-optimized homodyne detection of single-nanotube absorption**. (**a**) Electron diffraction patterns (left half: experimental, right half: simulated) uniquely determine the chiral index of this nanotube as (24,24). The black diagonal feature in the experimental pattern is from the blocking stick inside the TEM for dark-field imaging. (**b**) Homodyne modulation signal (|ΔI/I|) at various values of δ as in Fig. 1a. With δ decreasing from 5 to 0.5 degrees, the signal at resonances gradually increases to 1% level. (**c**) Modulation signal at δ = ±0.04. It shows interference signature that when δ crosses zero, local

oscillator $E_{LO}$ changes sign and so does the homodyne interference modulation $\Delta I/I$. (**d**) The absolute absorption cross-section per carbon atom with both parallel (//) and perpendicular ($\perp$) light polarization. The $\sigma_{//}$ spectrum shows clear resonance peaks corresponding to the exciton transitions; while $\sigma_\perp$ is mostly featureless. For all studied SWNTs, $\sigma_\perp$ has an integrated absorption between 1/5 and 1/3 of that in $\sigma_{//}$ due to depolarization effect in 1D nanotubes.

**Figure 3**. **Representative absorption cross-section spectra of four nanotubes.** (**a**) semiconducting (18,14). (**b**) semiconducting (25,23). (**c**) metallic (16,16). (**d**) metallic (24,24). Black, red, violet and dark yellow lines are experimental data, phenomenological fitted total absorption, exciton component, and continuum component, respectively. The integer number p at each peak indexes the optical transition type as described in the text. The resonant peaks show symmetric shape and reveal the exciton nature of nanotube optical transitions; while an obvious continuum background exists. The phenomenological model described in the text reproduces the absorption spectra nicely. The exciton transition energy, oscillator strengths, and linewidth vary significantly with the nanotube species and transition index p.

**Figure 4**. **Tube-dependent exciton oscillator strength and transition linewidth in 57 SWNTs.** The diamonds and the circles represent data from 18 metallic and 39 semiconducting nanotubes, respectively. (**a**) Integrated absorption cross-section in the energy range of 1.55 to 2.48 eV. All the data converge to the value of graphene integrated over the same energy region (dashed line). This convergence originates from an approximated f-sum rule. (**b**) Tube-dependent exciton oscillator strength is described by a universal scaling law of $4.6\times10^{-24}/[(p+7.5)\cdot d]$ eVcm$^3$/atom (dashed line), where d is nanotube diameter and p is transition order indexing both semiconducting and metallic nanotube optical transitions. Metallic and semiconducting nanotubes, surprisingly, exhibit similar exciton oscillator strength, although they are characterized by significantly different electron-hole interactions. (**c**) Exciton transition linewidth in different nanotubes. In average the linewidth increases linearly with the transition energies. The higher slope for metallic nanotube (long-dashed line) than that for semiconducting nanotube (short-dashed line) reveals (in average) slightly shorter exciton lifetime due to coupling to free electrons in metallic nanotubes. The large scattering present in the exciton linewidth suggests that ultrafast dynamics of the excited states can vary dramatically with the exact nanotube electronic structure.

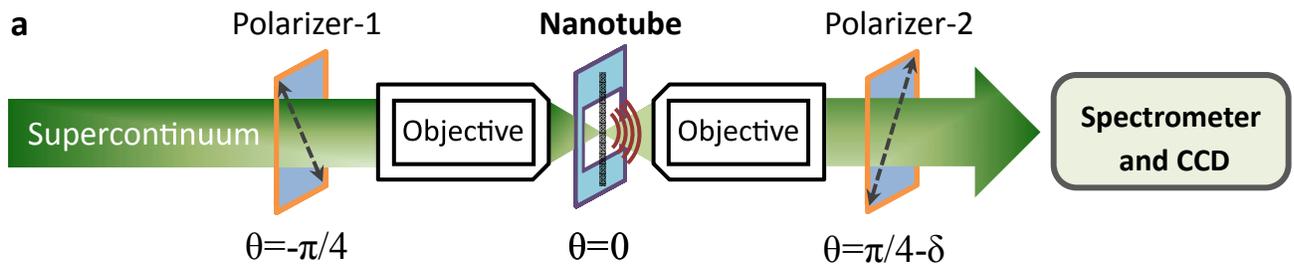
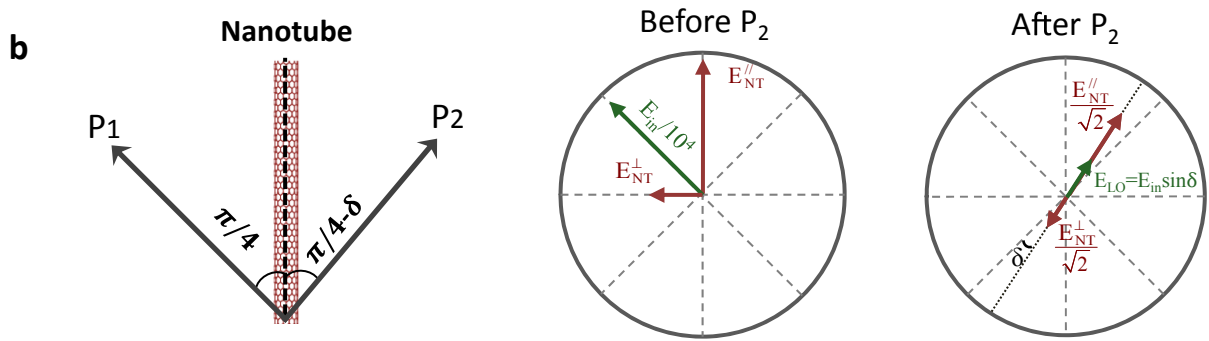

**Figure 1**

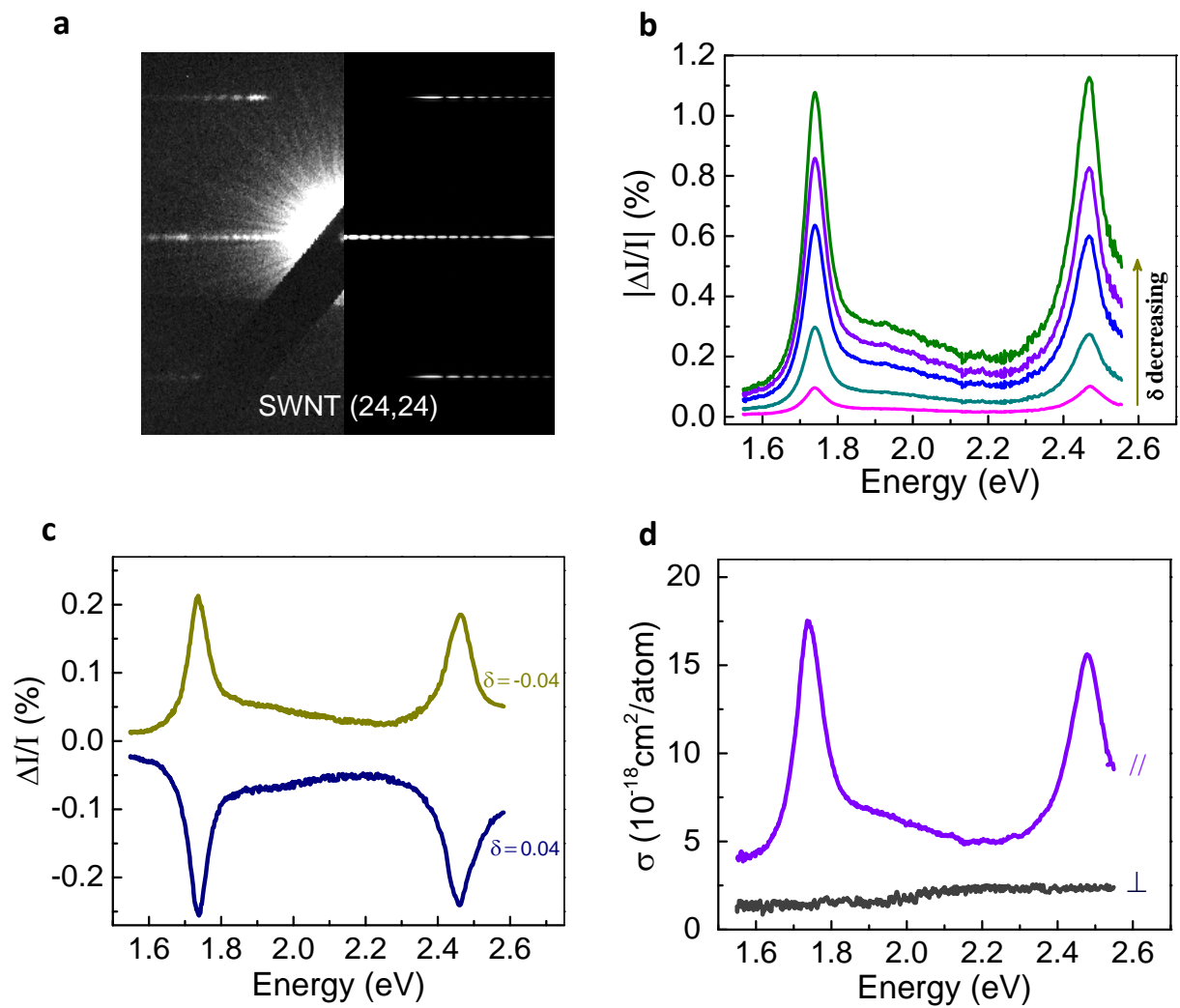

**Figure 2**

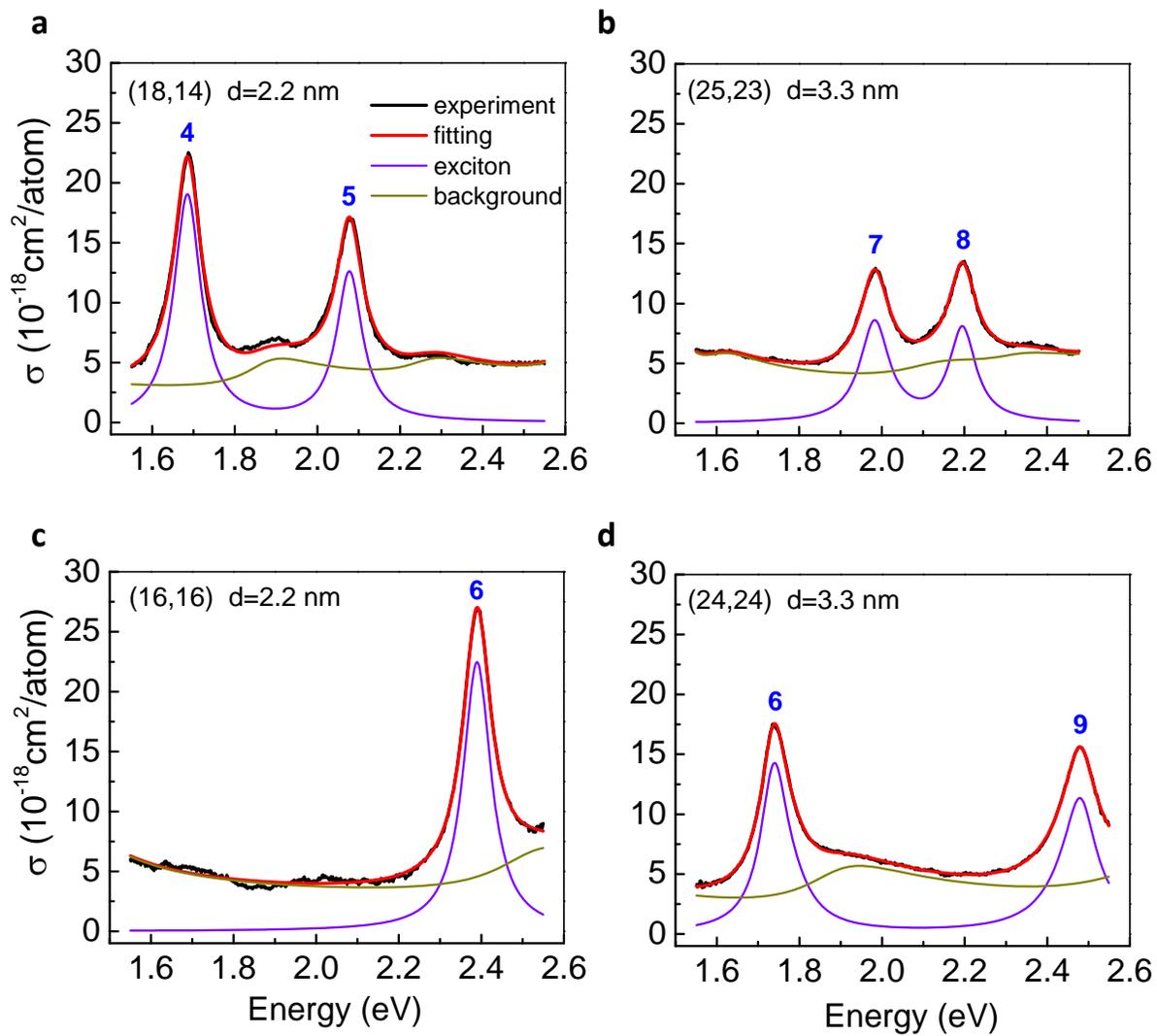

**Figure 3**

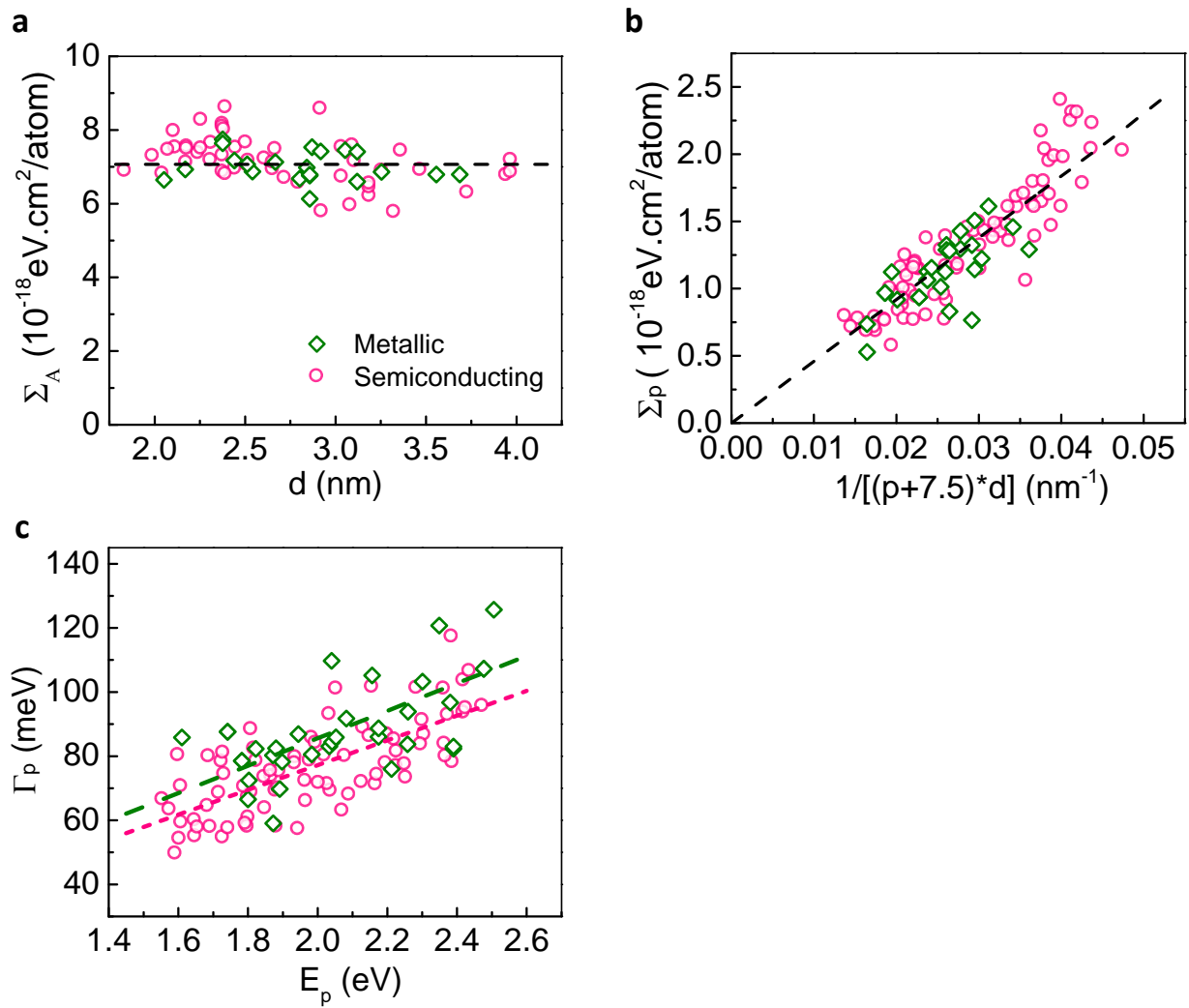

**Figure 4**

Supplementary Information for

# Systematic Determination of Absolute Absorption Cross-section of Individual Carbon Nanotubes

Kaihui Liu*[1,2], Xiaoping Hong*[1], Sangkook Choi[1,3], Chenhao, Jin[1], Rodrigo B. Capaz[1,4], Jihoon Kim[1], Wenlong Wang[2], Xuedong Bai[2], Steven G. Louie[1,3], Enge Wang[5], Feng Wang[1,3]

**S1. Schematic illustration of single-tube absorption and TEM characterization**

**S2. Determine absorption with both parallel and perpendicular polarizations**

**S3. Determine the absolute absorption cross-section from absorption constants**

**S4. Empirical formula for absorption cross-section of any (n,m) nanotube**

**S1. Determine the chiral index and absorption spectrum of the same individual carbon nanotube.**

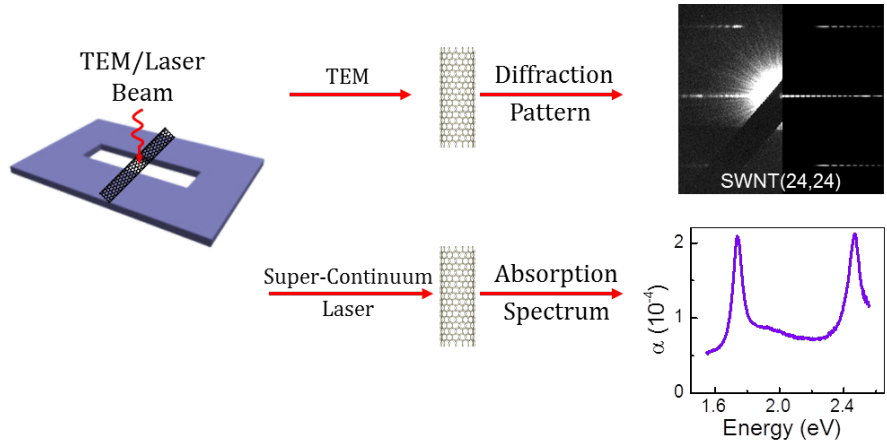

Figure S1. Scheme of combined single-nanotube TEM electron diffraction and absorption techniques

The Si/SiO$_2$ substrate is etched with open slit, on top of which suspended nanotubes are directly grown. Transmission electron microscope (TEM) beam and laser beam can both go through the slit. This design enables the combination of TEM electron diffraction and optical absorption spectroscopy techniques to investigate the chiral index and absorption cross-section of the same individual suspended carbon nanotubes.

**S2. Determine the nanotube absorption constants $α_{//}$ and $α_⊥$ from the homodyne modulation signal.**

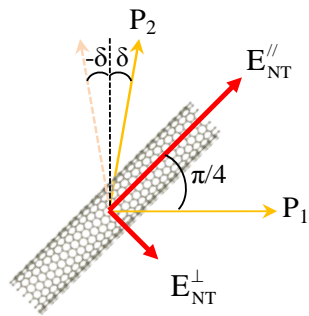

Figure S2. Geometric setting of two polarizers and the nanotube

In our experiment, we fix the relative angle between the first polarizer (P$_1$) to the nanotube axis as π/4. The second polarizer (P$_2$) is placed close to π/2 relative to the first polarizer with small deviation angle δ. The intensity modulation is characterized by absorption constants α$_{//}$ and α$_\perp$ for light polarized parallel and perpendicular to the nanotube axis, respectively. As described in the text, α$_{//}$ and α$_\perp$ are

$$\alpha_{//} = -\frac{2|E_{NT}^{//}|}{|E_{in}^{//}|}\cos\phi_{//} = \frac{-2|E_{NT}^{//}|}{|E_{in}|\cos(\frac{\pi}{4})}\cos\phi_{//},$$

$$\alpha_{\perp} = -\frac{2|E_{NT}^{\perp}|}{|E_{in}^{\perp}|}\cos\phi_{\perp} = \frac{-2|E_{NT}^{\perp}|}{|E_{in}|\cos(\frac{\pi}{4})}\cos\phi_{\perp}.$$

(Eq. S1)

Here E$_{in}$ is the incident electric field after the first polarizer; $E_{NT}^{//}$ ($E_{NT}^{\perp}$), $E_{in}^{//}$ ($E_{in}^{\perp}$) are nanotube scattered field and the incident field polarized in the parallel (perpendicular) direction of the nanotube axis, respectively; $\phi_{//}$ ($\phi_\perp$) is the relative phase between $E_{NT}^{//}$ ($E_{NT}^{\perp}$) and $E_{in}^{//}$ ($E_{in}^{\perp}$); the $\cos(\frac{\pi}{4})$ term comes from $\frac{\pi}{4}$ angle between the first polarizer and nanotube axis.

As for homodyne detection, $E_{NT}^{//}$ and $E_{NT}^{\perp}$ will combine at the second polarizer with their respective projection ratio $\cos(\frac{\pi}{4}-\delta)$ and $\cos(\frac{3\pi}{4}-\delta)$; meanwhile incident electric field E$_{in}$ will be converted to $E_{LO} = E_{in}\sin(\delta)$ by the second polarizer. Therefore the homodyne modulation signal (ΔI/I) is

$$-\frac{\Delta I}{I}(\delta) = \frac{-2\left[|E_{NT}^{//}|\cos(\frac{\pi}{4}-\delta)\cos\phi_{//} + |E_{NT}^{\perp}|\cos(\frac{3\pi}{4}-\delta)\cos\phi_{\perp}\right]}{|E_{in}|\sin(\delta)}$$

$$= \frac{1}{\sin(\delta)}[\alpha_{//}\cos(\frac{\pi}{4})\cos(\frac{\pi}{4}-\delta) + \alpha_{\perp}\cos(\frac{\pi}{4})\cos(\frac{3\pi}{4}-\delta)] \quad \text{(Eq. S2)}$$

$$= \frac{\alpha_{//}-\alpha_{\perp}}{2}\operatorname{ctg}(\delta) + \frac{\alpha_{//}+\alpha_{\perp}}{2}.$$

In principle, $\alpha_{//}$ and $\alpha_{\perp}$ can be extracted from any two intensity modulation signal taken at two different $\delta$ angles. For example, if we use $\pm\delta$ to yield two modulation signals $\Delta I/I(\pm\delta)$, the absorption constants can be expressed as

$$\alpha_{//} = \frac{\Delta I/I(-\delta) - \Delta I/I(\delta)}{2}\tan(\delta) - \frac{\Delta I/I(\delta) + \Delta I/I(-\delta)}{2},$$
$$\alpha_{\perp} = \frac{\Delta I/I(\delta) - \Delta I/I(-\delta)}{2}\tan(\delta) - \frac{\Delta I/I(\delta) + \Delta I/I(-\delta)}{2}. \quad \text{(Eq. S3)}$$

In our experiment, we use several pairs of $\Delta I/I(\pm\delta)$ and obtained the absorption constants $\alpha_{//}$ and $\alpha_{\perp}$ through fitting. One example of $\alpha_{//}$ and $\alpha_{\perp}$ for (24,24) SWNT is shown in Fig. S3.

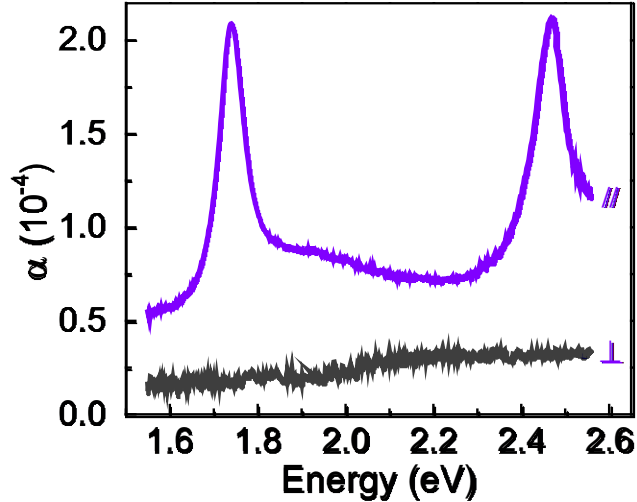

Figure S3. Absorption of (24-24) SWNT with parallel and perpendicular polarization

## S3. Determine the absolute absorption cross-section from absorption constants

The spatial profile of the supercontinnum at the focus has a Gaussian form and can be describe as

$$E = E_0 \cdot \exp\left(-\frac{(x-x_c)^2 + (y-y_c)^2}{R^2}\right), \quad \text{(Eq. S4)}$$

where $x_c$ and $y_c$ are the coordinates of the center position of the focus and R is a measure of the beam size. For a one-dimensional (1D) nanotube along y-direction and positioned at x, the absorption signal integrated over the nanotube length is

$$\alpha = \sigma \frac{\sqrt{2\pi}}{S_0} \cdot \frac{d}{R} \cdot \exp\left(-\frac{2(x-x_c)^2}{R^2}\right), \quad \text{(Eq. S5)}$$

where $\sigma$ is the absorption cross-section per atom, $S_0$ is the area per carbon atom in graphitic lattice and d is the nanotube diameter. We systematically measured the absorption spectra with the nanotube at different positions in the focused beam. Figure S4 shows the absorption signal at 710 nm, from which we can determine the beam radius R and center position $x_c$ by fitting. We performed the fitting for all wavelengths in our spectral range. With the beam size information, we obtain absolute values of the nanotube absorption cross-section in the whole spectral range.

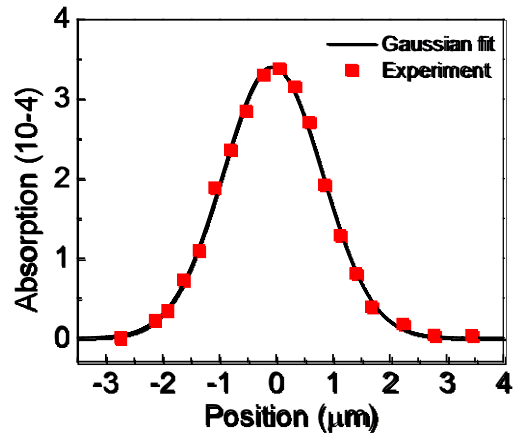

Figure S4. Dependence of absorption at 710 nm on nanotube position in the beam focus and its fitting to Gaussian function of Eq. S5

## S4. Empirical formula for absorption cross-section of an (n,m) nanotube

We developed an empirical description of the absorption cross-section of an (n,m) nanotube based on the experimentally extracted parameters. Each transition will contribute a Lorentzian peak and a tail at higher energy side described by

$$\frac{\Sigma_p}{\pi} \cdot \frac{w_p}{(E-E_p)^2+w_p^2} + \frac{\Sigma_p}{a \cdot \pi} \cdot \text{conv}\left(\frac{b \cdot w_p}{E^2+(b \cdot w_p)^2}, \frac{\Theta[E-(E_p+\Delta)]}{\sqrt{E-(E_{ii}+\Delta)}}\right), \quad \text{(Eq. S6)}$$

where $\Theta$ is the heavyside step function, conv is the convolution operation, $E_p$, $\Sigma_p$, $w_p$ are respectively the excitonic peak energy, oscillator strength and half linewidth for each optical transition. a, b and $\Delta$ describe, respectively, the oscillator strength, energy broadening, and energy offset of the higher energy tail relative to the exciton transition. They are set as constants for all transitions in a given carbon nanotube. p is an integer indexing optical transitions of both semiconducting ($S_{ii}$) and metallic nanotubes ($M_{ii}$) starting from 1 in the order of $S_{11}$, $S_{22}$, $M_{11}$, $S_{33}$, $S_{44}$, $M_{22}$, $S_{55}$, $S_{66}$, $M_{33}$, $S_{77}$,…. Sum of contributions (value of Eq. S6) from all resonances at a particular energy gives the total absorption cross-section at that energy. Resonance peak positions are known from the nanotube optical transition atlas established by Raleigh scattering spectroscopy in our recent work[1]. The oscillator strength and linewidth follow empirical scaling laws described in the text. Parameters a, b and $\Delta$ are also approximated empirically as linear functions of nanotube diameter. All the parameters are summarized in Table S1. This phenomenological description reproduces the absorption spectra within 20% accuracy in our experimental energy region 1.45-2.55 eV. One representative comparison to the experimental data is shown in Fig. S5.

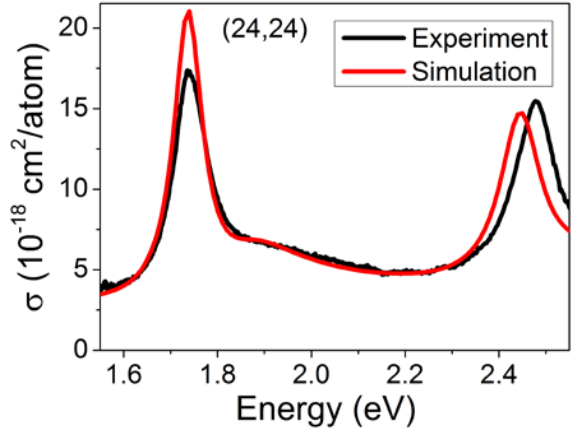

Figure S5 Experimental data (black) and empirical prediction of absorption cross-section (red) of (18,14) nanotube

Table S1. Parameters for the empirical description of absorption cross-section. $\Sigma_p$ is in unit of $10^{-18}$ eV.cm$^2$/atom; $w_p$ and $\Delta$ are in unit of eV; a and b are dimension-less numbers; d is in unit of nm. p is an integer indexing optical transitions of both semiconducting ($S_{ii}$) and metallic nanotubes ($M_{ii}$) starting from 1 in the order of $S_{11}$, $S_{22}$, $M_{11}$, $S_{33}$, $S_{44}$, $M_{22}$, $S_{55}$, $S_{66}$, $M_{33}$, $S_{77}$,…

|  | Semiconducting | Metallic |
|---|---|---|
| $\Sigma_p$ | 45.9/[(p+7.5)d] | 45.9/[(p+7.5)d] |
| $w_p$ | 0.0194*$E_p$ | 0.0214*$E_p$ |
| a | 4.673-0.747*d | 0.976+0.186*d |
| b | 0.97+0.256*d | 3.065-0.257*d |
| $\Delta$ | 0.273-0.041*d | 0.175-0.0147*d |